\documentclass{article}

\usepackage{PRIMEarxiv}

\usepackage[utf8]{inputenc} 
\usepackage[T1]{fontenc}    
\usepackage{hyperref}       
\usepackage{url}            
\usepackage{booktabs}       
\usepackage{amsfonts}       
\usepackage{nicefrac}       
\usepackage{microtype}      
\usepackage{lipsum}
\usepackage{fancyhdr}       
\usepackage{graphicx}       
\usepackage[authoryear]{natbib}

\title{A weighted-likelihood framework for class imbalance in Bayesian prediction models
}

\author{
  Stanley E. Lazic \\
  Prioris.ai Inc. \\
  Ottawa, Canada \\
  \texttt{stan.lazic@cantab.net}
}

\begin{document}
\maketitle

\begin{abstract}
Class imbalance is a pervasive problem in predictive toxicology, where the number of non-toxic compounds often exceeds the number of toxic ones. Models trained on such data often perform well on the majority class but poorly on the minority class, which is most relevant for safety assessment. We propose a simple and general Bayesian framework that addresses class imbalance by modifying the likelihood function. Each observation's likelihood is raised to a power inversely proportional to its class proportion, with the weights normalized to preserve the overall information content. This weighted-likelihood (or power-likelihood) approach embeds cost-sensitive learning directly into Bayesian updating. The method is demonstrated using simulated binary data and an ordered logistic model for drug-induced liver injury (DILI). Weighting alters parameter estimates and decision boundaries, improving balanced accuracy and sensitivity for the minority (toxic) class. The approach can be implemented with minimal changes in standard probabilistic programming languages such as {\tt Stan}, {\tt PyMC}, and {\tt Turing.jl}. This framework provides an easily extensible foundation for developing Bayesian prediction models that better reflect the asymmetric costs of safety-critical decisions.
\end{abstract}


\section{Introduction}
Class imbalance is common in toxicological datasets, where the number of compounds classified as non-toxic typically exceeds those classified as toxic. As a result, predictive models often achieve deceptively high overall accuracy while performing poorly on the minority (toxic) class, which is the outcome of greatest biological and regulatory concern. Standard performance metrics can therefore give an inflated impression of model quality, especially when the goal is to detect rare but important safety signals.

Existing solutions to class imbalance can be grouped into two broad categories\citep{Krawczyk2016}. Sampling-based methods modify the data directly through oversampling, undersampling, or synthetic data generation (e.g., SMOTE, ADASYN; \cite{Chawla2002,He2008}). Cost-sensitive methods instead modify the learning objective so that minority-class observations contribute more strongly to the result. For Bayesian models, a recent alternative by \citet{Nassiri2024} adjusts posterior class probabilities to reflect class proportions in the training data, but leaves parameter estimates, decision boundaries, and standard classification metrics unchanged. In many toxicological applications, however, it is desirable to adjust these internal components as well, so that predictive uncertainty and parameter inference both reflect the true cost asymmetry between errors.
 
Here, we describe how to account for class imbalance by modifying the loss function, which is the log-likelihood of a Bayesian prediction model. This is consistent with the approach taken in many popular frequentist machine learning models and is easy to implement within a Bayesian framework. Suppose we have $N$ independent observations $x$, a prior distribution $p(\theta)$ describing our uncertainty about the parameters before seeing the data, and a likelihood function $p(x | \theta)$ describing how the data arise from the model. Bayesian updating then combines these to obtain the posterior distribution $p(\theta | x)$, which represents our updated uncertainty after observing the data. Information on class imbalance can be incorporated by raising each observation's likelihood contribution to a power ($w_i$), giving

\begin{equation}
  p(\theta | x_{1:N}) \propto p(\theta) \prod_{i=1}^{N}  p(x_i | \theta)^{w_i},
\end{equation}
\noindent where $w_i > 0$ and $\sum_{i} w_i = N$.

Power-likelihoods are not a new idea in Bayesian inference. This approach is used to incorporate and downweight historical data \citep{Chen2000,Ibrahim2003}, account for misspecified likelihoods \citep{Holmes2017,Aitchison2021}, make the model robust to outliers \citep{Ghosh2015}, update distributions over parameters with a general loss function instead of a traditional likelihood function \citep{Bissiri2016,Lyddon2019}, and even to account for biased judgements in Bayesian models of human decision making \citep{Matsumori2018}. Research, and debate, in this area often focuses on how to determine suitable values for $w$, and depending on the model, $w$ could be a scalar or a vector of length $N$. When accounting for class imbalance, each sample gets its own weight, which is objectively set by the inverse proportion of its class size, and with the restriction that the weights sum to the sample size ($N$).

This simple approach removes the bias due to class imbalance from the start, rather than counteracting it as a post-processing step. Modifying the likelihood will result in different parameter estimates, and therefore different predictions. This approach has the advantage of being applicable to general cost-sensitive predictions where errors have different costs. By incorporating class imbalance directly into Bayesian updating, the proposed framework prioritizes detection of rare but toxic compounds, a common challenge in predictive toxicology where false negatives carry significant safety, regulatory, and financial consequences.

\section{Methods}

\subsection{Data}
Two datasets will be used to demonstrate the effects of accounting for class imbalance. The first is simulated data for a binary outcome with a large class imbalance. The class proportions are Class 0 = 0.87 and Class 1 = 0.13. The data consists of 100 samples with two continuous predictor variables.

The second data set is the drug-induced liver injury (DILI) data from \cite{Williams2020}. The data includes 96 compounds and the outcome is an ordered three-level categorical variable, where the categories represent increasing levels of liver toxicity.  Class 1 represents the safest compounds, Class 2 has mild hepatotoxicity, and Class 3 are the most toxic. Class proportions are 0.34, 0.42, and 0.24, respectively. The predictors are the results from several \textit{in vitro} toxicity assays, the degree to which a compound favours a lipid versus an aqueous environment, and the maximum clinical concentration in the blood (see the original paper for details).

\subsection{Models}
A Bayesian binary classification model can be represented as

\begin{eqnarray*}
  y_i & \sim & \mathrm{Bernoulli}(\eta_i)^{w_i} \quad i = 1\dots N\\
  \mathrm{logit}(\eta_i) & = & X_{ij}\beta_j \\
  \beta_j & = & \mathrm{Normal}(0, \sigma)
\end{eqnarray*}

\noindent where $y$ is a vector of $N$ of binary values indicating the class, and $i$ indexes the sample (e.g. patient or compound). $\eta$ is the linear predictor, defined by an $i$-by-$j$ matrix of predictor variables ($X$) and the model parameters $\beta$ estimated from the data. A Bayesian model also includes a prior over the $\beta$ parameters, which is given here as a Normal distribution with hyperparameters mean 0 and standard deviation $\sigma$. The likelihood can be modified to account for the class imbalance by raising it to a power that is inversely proportional to the proportion of samples in each class ($w$). Each sample $i$ gets a weight that is the same for all samples in that class. It is crucial that the weights sum to the sample size, $N$, otherwise the amount of information in the data will be artificially increased or decreased.

For example, if 75\% of the samples are Class 0 and 25\% are Class 1, the inverse proportions are 1/0.75 = 1.33 and 1/0.25 = 4. Each sample is assigned its respective value, and the weights are then normalised by dividing by their sum and multiplying by $N$. If there are 8 samples, the unnormalised weights are \{1.33, 1.33, 1.33, 1.33, 1.33, 1.33, 4, 4\}, the sum of the weights is 16, which gives normalised weights of \{0.67, 0.67, 0.67, 0.67, 0.67, 0.67, 2, 2\}. The Github repository provides an R function to perform these calculations for two or more classes.

Incorporating weights into models defined using standard Bayesian software is a straightforward one- or two-line modification. In {\tt Stan} \citep{Carpenter2017}, the line of code that increments the log-likelihood is multiplied by the weights, which are passed in as data. This operation can be vectorised so that {\tt y}, {\tt eta} (on the probability scale), and {\tt w} are vectors. 

\begin{verbatim}
  target += bernoulli_lpmf(y | eta) * w; 
\end{verbatim}

When using the {\tt Turing.jl} package in Julia \citep{Ge2018}, the {\tt @addlogprob!} macro enables the log-likelihood to be multiplied by the weights in a similar manner, where {\tt eta} is on the probability scale.

\begin{verbatim}
  @addlogprob! sum(loglikelihood.(Bernoulli.(eta), y) .* w)
\end{verbatim}

When using the {\tt PyMC} Python package \citep{AbrilPla2023}, the {\tt Potential()} method enables the weights to be incorporated, where {\tt eta} once again is on the probability scale.

\begin{verbatim}
  logprob = pm.Bernoulli.logp(y, p=eta)
  pm.Potential("weighted_LL", (w * logprob).sum())
\end{verbatim}

An unweighted analysis can be performed in all three languages by supplying a vector of 1's for the weights. Hence, the same code and model definitions can be used for both weighted and unweighted analyses. The weighted likelihood can also be specified on the log-odds scale for {\tt eta} instead of the probability scale in the above models. The Github repository provides full model specifications for each language, and the results presented here are based on the {\tt Stan} model. Extensive simulations comparing unweighted and weighted models have not been conducted because one method is not uniformly better; some metrics will improve while others will deteriorate, and the best method will need to be determined empirically for each application.

The DILI data model is identical to that described in \cite{Williams2020}, except for using a weighted log-likelihood. This model is only implemented in {\tt Stan}. The predictions and reported metrics are based on leave-one-out (LOO) validation. Example models are implemented in {\tt Stan}, {\tt PyMC}, and {\tt Turing.jl}, and all code and reproducible scripts are archived on Github (\url{https://github.com/stanlazic/weighted_likelihoods}).

\section{Results}

\subsection{Simulated binary data}

The simulated data has two predictors: $X1$, where higher values are associated with Class 1 (Fig. 1A), and $X2$, where higher values are associated with Class 0 (Fig. 1B). Weighted and unweighted models yield different estimates of the intercept ($\beta_0$; Fig. 1C), resulting in different decision boundaries (Fig. 1D) and therefore different predictions. Figure 1E shows the median posterior predicted probability for each compound in the unweighted analysis. Since the predictions are heavily influenced by samples from Class 0, the probabilities tend to be low. A weighted analysis that accounts for class imbalance has higher predicted probabilities for the Class 1 samples, and the predictions for all samples cover most of the probability range (Fig. 1F). The horizontal dashed lines in Figure 1E and F represent a common decision threshold at $p = 0.5$, which classifies samples above the threshold as belonging to Class 1 and those below the threshold as belonging to Class 0. The weighted analysis better captures the true Class 1 samples, whereas the unweighted analysis is better at classifying true Class 0 samples. Is there a net benefit from weighted analysis?

\begin{figure}[htb]
\begin{center}
\includegraphics[scale=0.6]{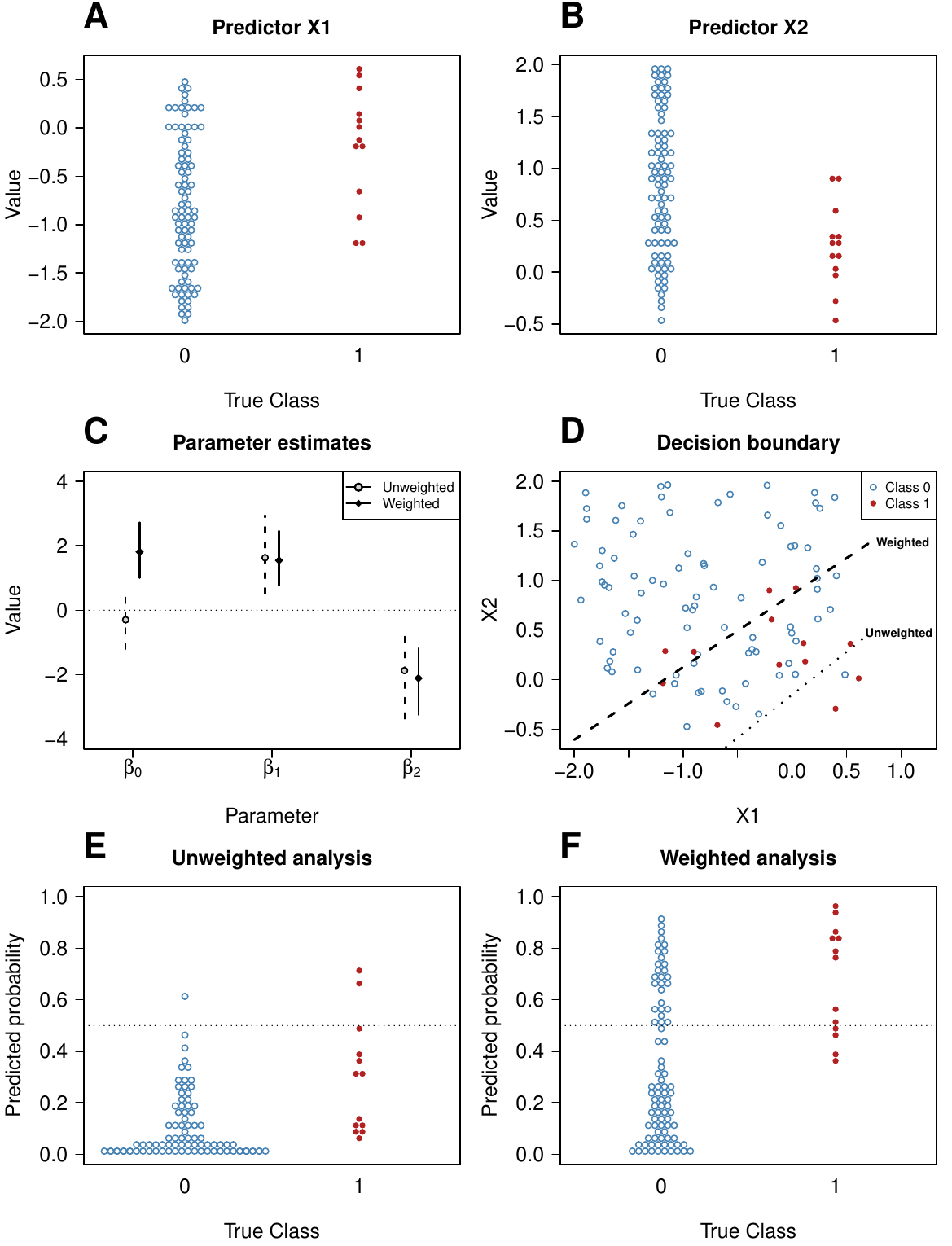}
\caption{Simulated data example with class imbalance. High values of predictor $X1$ are associated with Class 1 (A) and high values of predictor $X2$ are associated with Class 0 (B). Parameter estimates (C) and decision boundaries (D) differ between unweighted and weighted Bayesian logistic regression models. An unweighted model predicts low values for most Class 1 samples due to class imbalance bias (E). A weighted model that accounts for class imbalance predicts higher values for most Class 1 samples and the predictions span most of the probability range (F).}
\end{center}
\end{figure}

Table 1 shows the metrics for the two analyses. Since certain metrics are inversely related, some improve while others worsen; this is an inevitable trade-off when adjusting for class imbalance. There is no difference in the area under the receiver operating characteristic curve (AUC-ROC) between the two models. However, some metrics such as accuracy and the Brier score are strongly affected by class imbalance and therefore the ``balanced'' version of these metrics should be used \citep{Lazic2020}. The unweighted analysis has an accuracy of 0.88 but a balanced accuracy of only 0.57, indicating a poor model. In the weighted analysis, accuracy and balanced accuracy are similar: 0.68 and 0.69, respectively, indicating (1) the weighting removed the bias resulting from class imbalance, and (2) the weighted analysis was better than the unweighted analysis on the basis of balanced accuracy.

\begin{table}[htb]
\begin{center}
\caption{Simulated data metrics. The better result is in bold font.}
\begin{tabular}{lll}
\toprule
Metric & Unweighted analysis & Weighted analysis \\
\midrule
  AUC & 0.83 & 0.83 \\
  Accuracy  & 0.88 & 0.68\\
  Balanced accuracy & 0.57 & \textbf{0.69} \\
  Brier score & 0.09 & 0.18\\
  Balanced Brier score & 0.28 & \textbf{0.17} \\
  Sensitivity (Recall) & 0.15 & \textbf{0.69} \\
  Specificity & \textbf{0.99} & 0.68 \\
  PPV (Precision) & \textbf{0.67} & 0.24\\
  NPV & 0.89 & \textbf{0.94} \\
  F1 score & 0.25 & \textbf{0.36} \\
  P4 metric & 0.39 & \textbf{0.49} \\
  \midrule
  Mean calibration (MSE) & \textbf{0.02} & 0.11 \\
  Intercept & \textbf{0.07} & -2.00 \\
  Slope & \textbf{0.93} & 0.88 \\
\bottomrule
\end{tabular}
\newline AUC = Area under the ROC curve; MSE = Mean squared error; PPV = Positive predictive value; NPV = Negative predictive value
\end{center}
\end{table}

The pattern of results is similar for the Brier score (lower values are better) and the weighted model performs much better. Sensitivity and specificity are inversely related and the unweighted analysis has excellent specificity (0.99) but poor sensitivity (0.15). The weighted analysis better balances these metrics and has a specificity of 0.68 and a sensitivity of 0.69. The sensitivity and positive predictive value (PPV; also called recall) are also inversely related, and the F1-score is the harmonic mean of these two quantities which provides a one-number summary. The P4-metric is a new score that combines sensitivity, specificity, PPV, and NPV into one number, and provides a helpful summary of these standard metrics \citep{Sitarz2023}. The weighted analysis is considerably better based on both the F1 and P4 metrics.

Overall, the weighted analysis appears to be preferable based on the metrics that have been examined so far. However, another often neglected aspect of model assessment is calibration \citep{VanCalster2019}, which are the metrics below the horizontal line in Table 1. A model is calibrated if the predicted probabilities correspond to the observed event proportions. Calibration is particularly important for safety-related models because overconfident or poorly calibrated probability estimates can lead to inappropriate compound prioritization. \cite{VanCalster2016} describe four increasingly stringent calibration levels. The lowest level is \textit{mean calibration}, which is the proportion of samples classified as Class 1. A model is well calibrated if the predicted proportion is similar to the observed proportion. The true proportion of Class 1 samples is 0.13 in the simulated data. Although the unweighted analysis underestimates this value (0.03), it is closer than the weighted estimate (0.37). Table 1 presents the results of mean calibration as the mean squared error (MSE) between the observed and predicted class proportions.

The next level is \textit{weak calibration} which is characterised by the intercept and slope of a regression model predicting the observed proportion of classes from the predicted probabilities of Class 1. A perfectly calibrated model has an intercept of 0 and a slope of 1. Based on these calibration metrics, the unweighted model is better, especially for the intercept. Due to the small sample size in this example, more stringent calibration levels cannot be assessed \cite{VanCalster2016}. 

Whether a weighted or unweighted analysis is preferable for any given dataset is an empirical question and depends on which metrics are prioritised and the purpose of the model; for example, will the model be used to ``rule-out'' or ``rule-in'' the presence of a toxic compound \citep{Hand2024}?

It is possible to improve the unweighted analysis metrics by using a threshold other than 0.5 to classify the samples. However, the threshold should be pre-specified, and as it depends on the degree of class imbalance, the optimal threshold will differ between studies and will not be known before conducting the analysis \citep{Collins2024}. Thresholds defined after a model has been built are often unstable, even when the sample size is reasonably large \citep{Wynants2019,Gerds2021}. The weighted analysis has the advantage that 0.5 is often a suitable threshold, which eliminates the need to estimate a value.

\subsection{DILI ordinal data}

Since DILI data exhibit only a modest degree of class imbalance (class proportions are 0.34, 0.42, and 0.24), the effect of accounting for class imbalance is relatively small. Figure 1 shows the posterior predictive distribution medians for each compound for the unweighted (Fig. 1A) and weighted (Fig. 1B) analyses. The main effect of weighting is to shift the top cut-point between Class 2 and 3 compounds (horizontal dashed line) slightly downwards so that more Class 3 compounds are classified correctly. This also results in a greater number of Class 2 compounds being incorrectly classified as Class 3. For both analyses, the posterior medians and the precision of the posterior predictive distributions are similar (Fig. 2C and D).

\begin{figure}[htb]
\begin{center}
\includegraphics[scale=0.54]{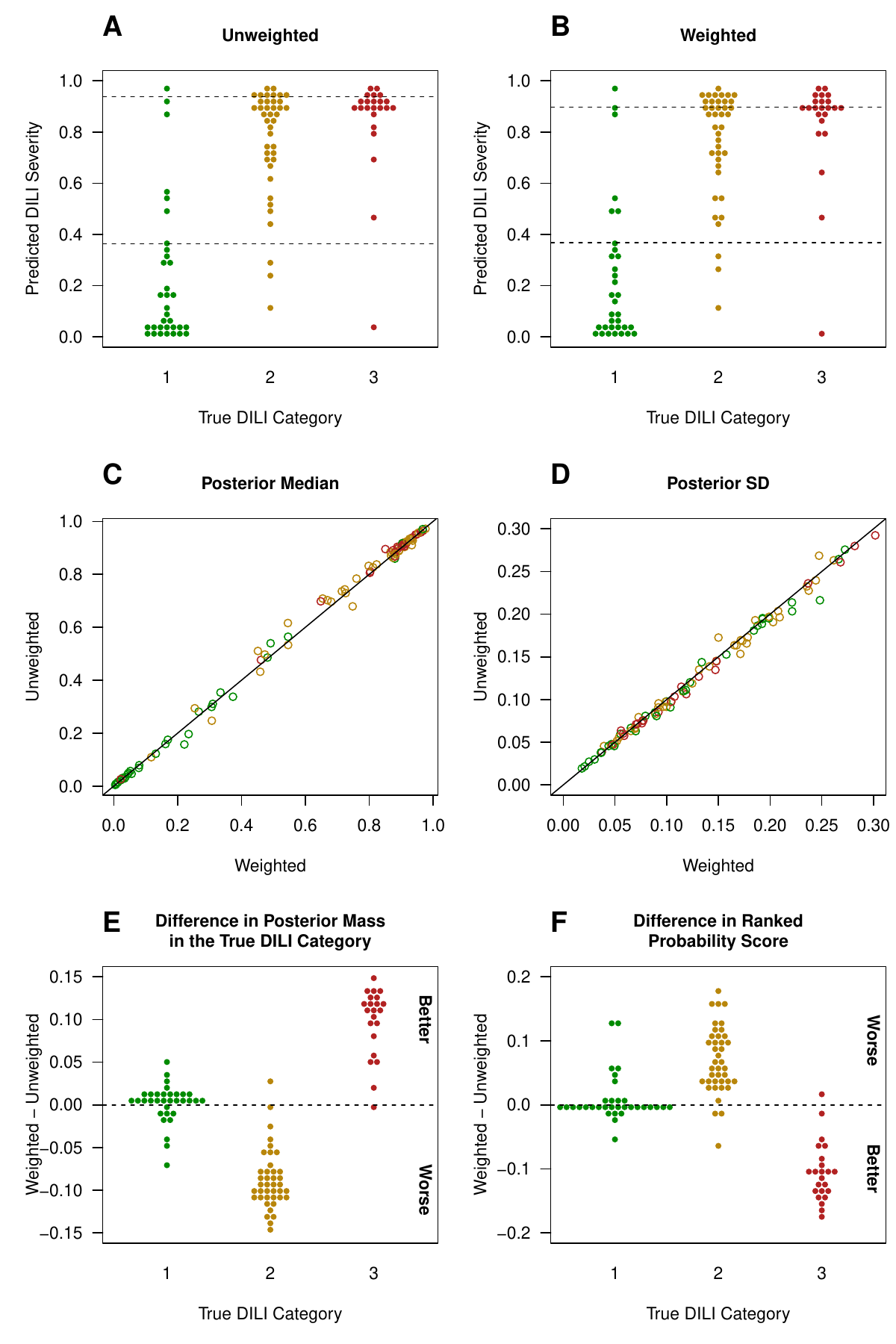}
\caption{Unweighted (A) and weighted (B) Bayesian ordinal regression model predictions for the DILI data. Both models return similar posterior median predictions for DILI severity (C) and uncertainty in the predictions (posterior standard deviations; D). The main effect of weighting is to lower the top cut-point that separates Class 2 and 3 compounds (top horizontal dashed line in A and B). Differences in the posterior mass of the true DILI category (E), and the ranked probability score (F) are measures of predictive accuracy and favour the weighted model for Category 3 compounds because these have a higher weight.}
\end{center}
\end{figure}

The greatest effect of weighting -- and the trade-offs being made -- are most clearly seen when calculating the proportion of the posterior predictive distribution that is in the true DILI category. Figure 2E displays this as the difference between the weighted and unweighted models. The weighting greatly improves the Class 3 compounds (minority class), but at the expense of the Class 2 compounds (majority class), with the Class 1 compounds having little change, on average. A similar pattern is observed for the ranked probability score, which is a generalisation of the Brier score to three or more ordered categories (Fig. 2F;  \citep{Murphy1970}). These results generalise to many toxicology datasets that have fewer samples for the toxic or most toxic class: weighting will improve a model's ability to detect the rare but toxic compounds.

The unweighted and weighted models have the same overall accuracy (0.70), but the weighted model has a slightly higher overall balanced accuracy (0.66 versus 0.65). The unweighted model is slightly better calibrated (mean calibration MSE: 0.03 vs. 0.05). Table 2 presents the binary metrics comparing adjacent DILI categories. 1 vs. \{2,3\} compares Category 1 compounds with the combination of Category 2 and 3 compounds, and \{1,2\} vs. 3 compares the combination of Category 1 and 2 compounds with Category 3 compounds. Based on the standard metrics, the unweighted model performs better when separating DILI Category 1 compounds from Category 2 and 3 compounds, but the weighted model performs slightly better when separating Category 3 compounds from Category 1 and 2 compounds. Once again, there are trade-offs to be made when using the weighted or unweighted model, and the preferred approach will depend on the purpose of the model. If the aim is to distinguish the most toxic Category 3 compounds from the other categories, then the weighted model is preferred (sensitivity = 0.52 versus 0.22).

\begin{table}[htb]
\begin{center}
\caption{DILI data metrics. The better result is in bold font when comparing methods within each DILI grouping.}
\begin{tabular}{lll|ll}
\toprule
  Metric & \multicolumn{2}{c}{1 vs. \{2,3\}} & \multicolumn{2}{c}{\{1,2\} vs. 3} \\
& Unweighted & Weighted & Unweighted & Weighted \\
\midrule
  AUC & 0.91 & 0.91 & \textbf{0.76} & 0.75 \\
  Accuracy  & \textbf{0.90} & 0.89 & 0.76 & 0.76 \\
  Balanced accuracy & \textbf{0.88} & 0.86 & \textbf{0.57} & 0.56 \\
  Sensitivity (Recall) & 0.94 & 0.94 & 0.22 & \textbf{0.52} \\
  Specificity & \textbf{0.82} & 0.79 & \textbf{0.93} & 0.79 \\
  PPV (Precision) & \textbf{0.91} & 0.89 & \textbf{0.50} & 0.44 \\
  NPV & 0.87 & 0.87 & 0.79 & \textbf{0.84} \\
  F1 score & \textbf{0.92} & 0.91 & 0.30 & \textbf{0.48} \\
  P4 metric & \textbf{0.88} & 0.87 & 0.45 & \textbf{0.60} \\
\bottomrule
\end{tabular}
\newline AUC = Area under the ROC curve; PPV = Positive predictive value; NPV = Negative predictive value
\end{center}
\end{table}

\section{Discussion}

Due to the inverse relationship between certain metrics, accounting for class imbalance will improve some metrics, but will often make others worse. In addition, accounting for class imbalance, using any method, can lead to miscalibrated predictions \citep{Gerds2021,Goorbergh2022,Piccininni2024}. This can be a serious problem if the imbalance in the training data reflects the expected imbalance in the population or environment where the model will be used. Adjustments for imbalance many improve metrics on the training data, but these models will perform poorly when used in production. Recalibration methods can potentially mitigate miscalibrated predictions, but they generally require larger sample sizes than the examples used here \citep{Kull2017,Guo2017}. For DILI prediction, weighting effectively lowers the threshold for classifying a compound as hepatotoxic, aligning with the precautionary principle typically applied in preclinical safety assessment. Thus, sensitivity is improved at the expense of specificity.

Sampling-based methods such as SMOTE are rarely appropriate for Bayesian analyses because they alter the effective sample size and artificially increase the apparent precision of posterior estimates. In contrast, the weighted-likelihood approach retains the true sample size while embedding cost sensitivity directly into the likelihood. Although some observations receive weights greater than one, normalization ensures that the total information content remains unchanged. In toxicological settings, where false negatives can have serious downstream consequences, this controlled re-weighting provides a principled alternative to ad hoc resampling.
 
Bayesian prediction models offer unique advantages for computational toxicology: they integrate prior biological knowledge, quantify multiple sources of uncertainty, and produce full posterior predictive distributions that can inform risk decisions \citep{Lazic2018,Reynolds2019a,Williams2020,Semenova2020,Reynolds2020,Lazic2021,Allen2022,Li2022,Spinu2022,Reinke2025}. Incorporating class imbalance through weighted likelihoods extends these benefits by aligning model training with toxicological priorities. The framework is simple to implement across major probabilistic programming platforms ({\tt Stan}, {\tt PyMC}, {\tt Turing.jl}) and directly supports reproducible, transparent workflows. Future work could extend this approach to multi-endpoint toxicity prediction, incorporate external biological priors, and evaluate calibration under large-scale imbalanced datasets such as Tox21 or eTOX.

\bibliography{/home/sel/areas/bib/BigBib}  
\bibliographystyle{abbrvnat}

\end{document}